\documentclass[graybox,natbib,nosecnum]{svmult}
\bibpunct{(}{)}{;}{a}{}{,} 

\pdfoutput=1   

\usepackage{mathptmx}       
\usepackage{helvet}         
\usepackage{courier}        
\usepackage{type1cm}        

\usepackage{makeidx}         
\usepackage{graphicx}        
\usepackage{multicol}        
\usepackage[bottom]{footmisc}
\usepackage[normalem]{ulem}	
\usepackage{hyperref}  
\usepackage[T1]{fontenc} 

\usepackage{soul}   

\usepackage{xspace, amssymb}

\makeindex             

\newcommand{\FeH}{\ensuremath{{\rm [Fe/H]}}\xspace}
\newcommand{\Kepler}{\textit{Kepler}\xspace}
\newcommand{\TESS}{\textit{TESS}\xspace}

\newcommand{\insitu}{\textit{In Situ}\xspace}
\newcommand{\migration}{\textit{Planet Migration}\xspace}
\newcommand{\pebbles}{\textit{Pebble Accretion}\xspace}

\begin{document}

\title*{Exoplanet Populations and their Dependence on Host Star Properties}

\author{Gijs D. Mulders}
\institute{Gijs D. Mulders 
\at Facultad de Ingenier\'ia y Ciencias, Universidad Adolfo Ib\'a\~nez, Av.\ Diagonal las Torres 2640, Pe\~nalol\'en, Santiago, Chile, \email{gijs.mulders@uai.cl} 
\at Millennium Institute for Astrophysics, Chile
}
%
%
\maketitle

\abstract{
Exoplanets around different types of stars provide a window into the diverse environments in which planets form.
This chapter describes the observed relations between exoplanet populations and stellar properties and how they connect to planet formation in protoplanetary disks.
Giant planets occur more frequently around more metal-rich and more massive stars. These findings support the core accretion theory of planet formation, in which the cores of giant planets form more rapidly in more solid-rich and more gas-rich protoplanetary disks. 
Smaller planets, those with sizes roughly between Earth and Neptune, exhibit different scaling relations with stellar properties. These planets orbit stars with a range of metallicities and occur more frequently around lower mass stars, indicating that planet formation takes place in a wide range of environments. 
Within M dwarfs, both radial velocity and transit surveys show that planets are smaller and located closer to the star when the stellar mass is lower. 
Additions to the core accretion model, in particular pebble accretion, have shown success in explaining the enhanced planet formation efficiency around low mass stars.
}

\section{Introduction}\label{s:intro}
\begin{figure}
\centering
\includegraphics[width=0.7\textwidth]{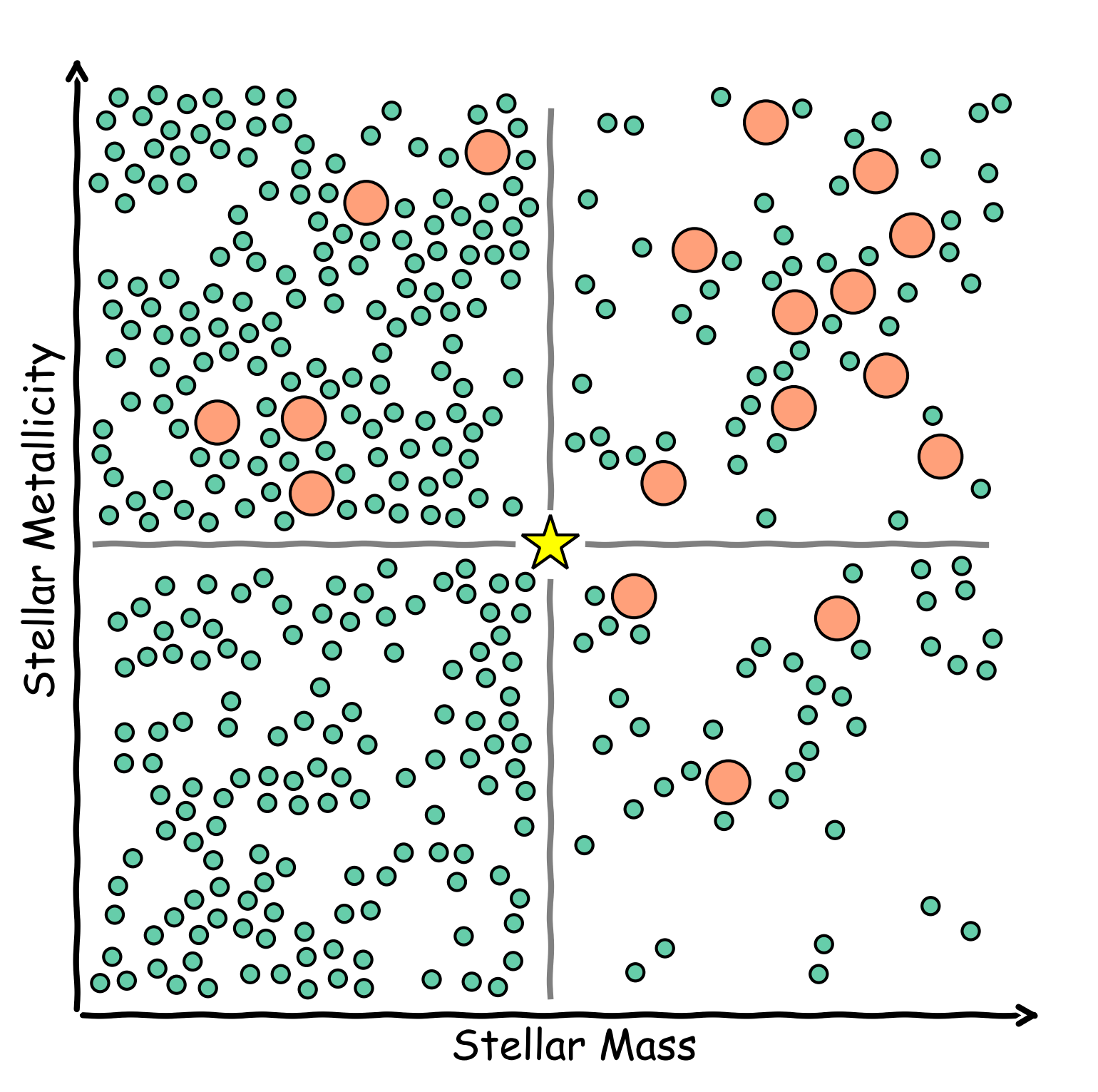}
\caption{Trends in the exoplanet population as function of stellar mass and metallicity, illustrating the different behavior of the giant planet population (large pink circles) and planets smaller than Neptune (small cyan circles). The location of the sun is indicated with a yellow star. The location of individual symbols is randomly generated, with the density of point corresponding to the exoplanet occurrence rate. 
Any resemblance between symbol locations and observed exoplanets is entirely coincidental.  
}
\label{f:MZ}
\end{figure}

Exoplanets are observed around a diverse set of host stars.
The first exoplanet discovered around a main-sequence star, 51 Pegasi b, orbits a star enriched in heavy elements (metals) compared to the sun \citep{1995Natur.378..355M}.
In contrast, one of the earliest discovered planets that could conceivably be rocky, Gliese 581e, orbits a metal-poor M dwarf less than a third the mass of the sun \citep{2009A&A...507..487M}.
While these discoveries represent just two examples of the more than five thousand exoplanets known to date\footnote{\url{https://exoplanetarchive.ipac.caltech.edu/}}, they are indicative of the broader trends between exoplanets and their host stars that have since emerged from exoplanet surveys, illustrated in Figure \ref{f:MZ}.
Giant planets occur more frequently around more massive and more metal-rich stars \citep[e.g.][]{2004A&A...415.1153S,2010PASP..122..905J,2018ApJ...860..109G}. Sub-Neptunes occur around stars with a wide range of metallicities \citep{2008A&A...487..373S,2012Natur.486..375B}, but occur more frequently around lower mass stars \citep{2012ApJS..201...15H,2015ApJ...798..112M,2021A&A...653A.114S,2023AJ....165..262Z}.

It is no coincidence that the smallest planets were first discovered around M dwarfs. The lower stellar mass compared to more sun-like stars with spectral types F, G and K facilitates the detection of less massive planets with radial velocity techniques \citep[e.g.][]{2003AJ....126.3099E}. Similarly, the small size of M dwarfs lead to deeper transits for a planet of the same size when compared to sun-like stars \citep{2013ApJ...767...95D}. 
Most exoplanets to date, however, have been discovered around F, G, and K dwarfs because more bright targets are observable.
Hence, an understanding of survey detection efficiency and selection biases are crucial to understand trends in the occurrence of the exoplanet population with host star properties. 

While the giant planet-metallicity correlation was initially interpreted as pollution of the stellar atmosphere by planetary systems \citep[e.g.][]{1997MNRAS.285..403G}, it is now widely accepted that the stellar metallicity is a proxy of the solid inventory of the protoplanetary disks in which planets form. It has been established that the increased occurrence of giant planets around high-metallicity stars arises because giant planet cores are more likely to form in disks with a larger amount of solids \citep[e.g.][]{2004ApJ...616..567I,2021A&A...656A..70E}. Similarly, the lower frequency of giant planets around M dwarfs can be explained by those stars having less massive disks \citep{2004ApJ...612L..73L,2021A&A...656A..72B}. 
The relation between exoplanets and their hosts stars provide stringent constraints on planet formation theory, as properties of exoplanet host stars trace the conditions in protoplanetary disks at the time of planet formation.

These trends, however, breaks down for planets smaller than Neptune, hereafter sub-Neptunes, which poses some urgent questions about the planet formation process.
Why is the frequency of sub-Neptunes almost independent of stellar metallicity, even when the initial inventory of condensible solids must have varied by an order of magnitude? 
Do the elevated planet occurrence rates around M dwarfs, where protoplanetary disk masses were lower, imply that there is something fundamentally different about the planet formation process around low-mass stars?

Not all stars are equally amenable for exoplanets discovery and certain types of stars have been more thoroughly searched than others.
To account for these selection and detection biases, planet occurrence rates can be calculated to infer trends in the intrinsic planet population. 
Variations in the planet occurrence rate with stellar parameters can be estimated from exoplanets surveys under the following conditions:
\begin{enumerate}
\item{The survey covers a range of stellar properties, with a sufficient number of planet detections across this range to identify trends.}
\item{Stellar properties are known for the surveyed stars, including those of stars without detected planets, to estimate the \textit{fraction} of stars with a given set of properties hosting planets. 
}
\item{The survey completeness can be estimated, to separate observation bias from intrinsic trends in the exoplanet population.}
\end{enumerate}
The focus of this chapter are trends identified in radial velocity and transit surveys with stellar mass and metallicity, which (mostly) satisfy these three requirements.
A notable omission are direct imaging surveys, though it should be noted that they are consistent with the positive stellar mass dependence of giant planets \citep[e.g.][]{2019AJ....158...13N,2021A&A...651A..72V}.

Trends for giant planets out to a few au and sub-Neptunes at orbital periods shorter than a few hundred days are discussed seperately. Emphasis is placed on studies that take into account the different observation bias and survey detection efficiency that exist when surveying planets around various type of stars.
These trends are then placed into the context of planet formation theory and models. An outlook for current and future surveys that can fill in some of the gaps in the current knowledge of the exoplanets populations around different types of stars is presented towards the end of this chapter, followed by a brief conclusion.

\section{Trends with Stellar Metallicity}\label{s:Z}
There is a general consensus that giant planet occurrence rates increase with host star metallicity, see also the review by \cite{2007ARA&A..45..397U}.  
The giant planet-metallicity relation is seen in radial velocity surveys of sun-like stars, M dwarfs, and evolved stars, and has also been identified for transiting planets.
However, Sub-Neptunes are found around stars with a wider range of metallicities, with no clear preference for metal-rich stars.
Throughout this chapter, the logarithm of the iron abundance with respect the solar abundance, \FeH, is used to represent stellar metallicity.

\begin{figure}
\includegraphics[width=\textwidth]{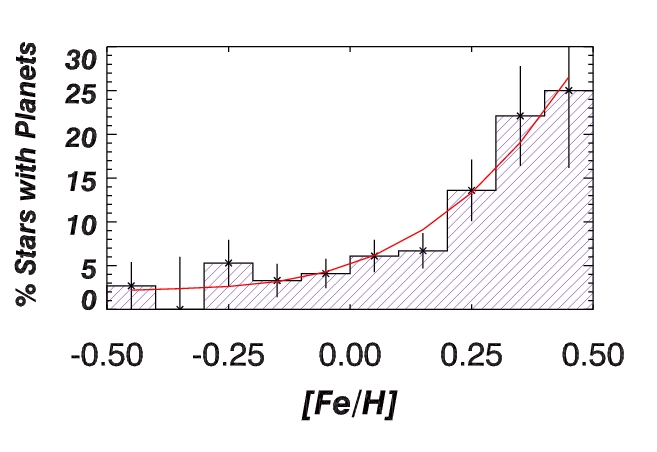} 
\caption{Giant planet occurrence rate as function of stellar metallicity, from \cite{2005ApJ...622.1102F} figure 5.
The red solid line shows a quadratic relation between planet occurrence and stellar metallicity ($\beta=2$, eq. \ref{eq:beta}).
Figure reproduced from \cite{2005ApJ...622.1102F} with permission from the authors.
}
\label{f:FeH:RV}
\end{figure}

\subsection{Positive Giant Planet-Metallicity Correlation}\label{s:Z:giants}
Giant planets occur more frequently around stars with higher metallicities (See Fig. \ref{f:FeH:RV}). 
Early indications of a planet-metallicity correlation were found by \cite{1997MNRAS.285..403G} based on metallicities of a handful of exoplanet hosts including 51 Peg b. The trend that giant planets are preferentially found around metal-rich host stars was subsequently recovered in larger samples \citep{1998A&A...334..221G,1998A&A...336..942F,2000A&A...363..228S,2001AJ....121..432G,2002PASJ...54..911S,2003AJ....125.2664L}. 

As outlined in the preceding section, characterizing star properties of non-planet hosts and detection efficiency of the survey are critical to separate observation bias from intrinsic planet population. \cite{2001A&A...373.1019S,2003A&A...398..363S} measured abundances for non-planet hosting stars 
and found that the giant planet hosts are systematically more metal-rich in a volume-limited sample. The detection frequency of giant planets was shown to increase with metallicity in a volume-limited sample of stars from the Hipparcos catalog \citep{2002PASP..114..306R}.
\cite{2004A&A...415.1153S} estimated planet occurrence rate as function of metallicity and identified a positive correlation at super-solar metallicities.
The high occurrence rate of giant planets around metal-rich stars was confirmed by (\citealt{2005ApJ...622.1102F}, see also Fig. \ref{f:FeH:RV}), who derived stellar abundances of stars in the Keck, Lick, and Anglo-Australian Telescope planet search surveys.

The occurrence rate of giant planets is a strong function of metallicity and scales roughly with the square of the number of iron atoms. At super-solar metallicities, $\FeH>0$, where planet detections are plenty, metallicity increases by a factor of 5 from $\sim 5\%$ at $\FeH=0$ to $\sim 25\%$ at $\FeH=0.5$. At lower metallicities, the shape of the metallicity distribution is less well quantified due to few planet detections, with a giant planet occurrence rate of approximately $\sim 2-3\%$ \citep[e.g.][]{2004A&A...415.1153S}. The functional form of the planet occurrence-metallicity correlation is often assumed to be a power-law\footnote{Note that \FeH is the logarithm of the iron abundance.}
\begin{equation}\label{eq:beta}
f_{\rm giant}\propto 10^{\beta \FeH},
\end{equation}
with index $\beta\approx2$ \citep[e.g.][]{2005ApJ...622.1102F,2007ARA&A..45..397U,2011A&A...533A.141S}.
\cite{2010PASP..122..905J} showed that such a functional form 
provides a better fit than a flat distribution at sub-solar metallicities. The planet occurrence rate likely continues to decrease at metallicities below $\FeH<-0.5$ \citep{2011A&A...526A.112S,2012A&A...543A..45M}, consistent with the non-detection of giant planets in metal-poor clusters and halo stars \citep[e.g.][]{2000ApJ...545L..47G,2021AJ....162...85B}. However, \cite{2013A&A...551A.112M} also argue that planet statistics at low metallicity are too small to discriminate between a linear function and a power-law. 

\runinhead{Transiting Giant Planets}
The planet-metallicity correlation has also been identified for transiting planets. The biggest challenge in identifying this correlation lies in characterizing stellar properties, in particular for non-planet host stars. The volume of stars searched for transiting planets is much larger than in radial velocity surveys -- both in terms of absolute numbers and galactic distance -- and characterization of stellar properties with high spectral resolution observations requires a significant investment in observing time. For this reason, most surveys have focused on characterizing planet-hosting stars \citep{2013ApJ...771..107E,2014Natur.509..593B,2017AJ....154..108J}.

A giant planet-metallicity relation in the \Kepler survey was first identified based on photometry by \cite{2011ApJ...738..177S}, who find that giant planet hosts have systematically redder colors than non-planet hosts, consistent with a metallicity increase in $0.2$ dex. Spectroscopic characterization of exoplanet host stars show that planets larger than $4 ~R_\oplus$ are preferentially found around stars with a super-solar metallicity of $0.15$--$0.18$ dex \citep{2012Natur.486..375B,2014Natur.509..593B,2017AJ....154...60W}. This result was confirmed by \cite{2013ApJ...771..107E} who found that giant planets only occur around high-metallicity stars ($\FeH >-0.05$ dex). 
\cite{2015AJ....149...14W} found 10 times more planets around metal-rich stars based on photometric metallicities, consistent with a power-law index $\beta=2$ as found in radial-velocity surveys. 

Occurrence rate calculations subsequently confirmed the giant planet metallicity relation, using both medium-resolution spectroscopic metallicities from \texttt{LAMOST} \citep{2016AJ....152..187M} and photometric metallicities for the non-planet hosts \citep[e.g.][]{2018AJ....155...89P}. The increased occurrence for giant planets has also been detected in ground-based transit data \citep{2020MNRAS.491.4481O}, while the hot Jupiter sample from the Transiting Exoplanet Survey Satellite, \TESS, is still under construction \citep[e.g.][]{2023ApJS..265....1Y}.

\runinhead{Dwarfs and Giants}
The giant planet-metallicity correlation is also present in stars with lower and higher masses than the sun.
Low-mass M dwarfs ($\lesssim 0.5 M_\odot$) are found to be enhanced in metallicity when they host giant planets \citep{2007A&A...474..293B,2009ApJ...699..933J,2012ApJ...748...93R,2012ApJ...747L..38T}. The exponent of the occurrence rate-metallicity correlation, in the range $\beta=[1.26,2.94]$, is consistent with that of sun-like stars \citep{2013A&A...551A..36N}.
The planet-metallicity correlation is less statistically robust than for FGK dwarfs due to a lower number of planet detections \citep{2010A&A...519A.105S,2014ApJ...791...54G}. 

Giant and sub-giant stars that have evolved off the main sequence provide an opportunity to measure planet occurrence rates around higher mass stars ($\gtrsim 1.5 M_\odot$).
The giant planet-metallicity correlation is less well established for these evolved stars than for main-sequence stars. \cite{2007A&A...475.1003H} found the first indications that evolved planet hosts are more metal-rich than non-planet hosts. Subsequent studies did often not find a planet-metallicity correlation \citep{2007A&A...473..979P,2008PASJ...60..781T,2013A&A...557A..70M}, showed mixed results \citep{2013A&A...554A..84M,2015A&A...574A..50J}, or did recover a correlation \citep{2017AJ....153...51W}.
Limiting this chapter to planet occurrence rate studies, i.e. those that take into account detection efficiency and sample selection, the planet occurrence rate is found to increase with stellar metallicity \citep{2010PASP..122..905J,2015A&A...574A.116R,2016A&A...590A..38J}.

\begin{figure}
\includegraphics[width=\textwidth]{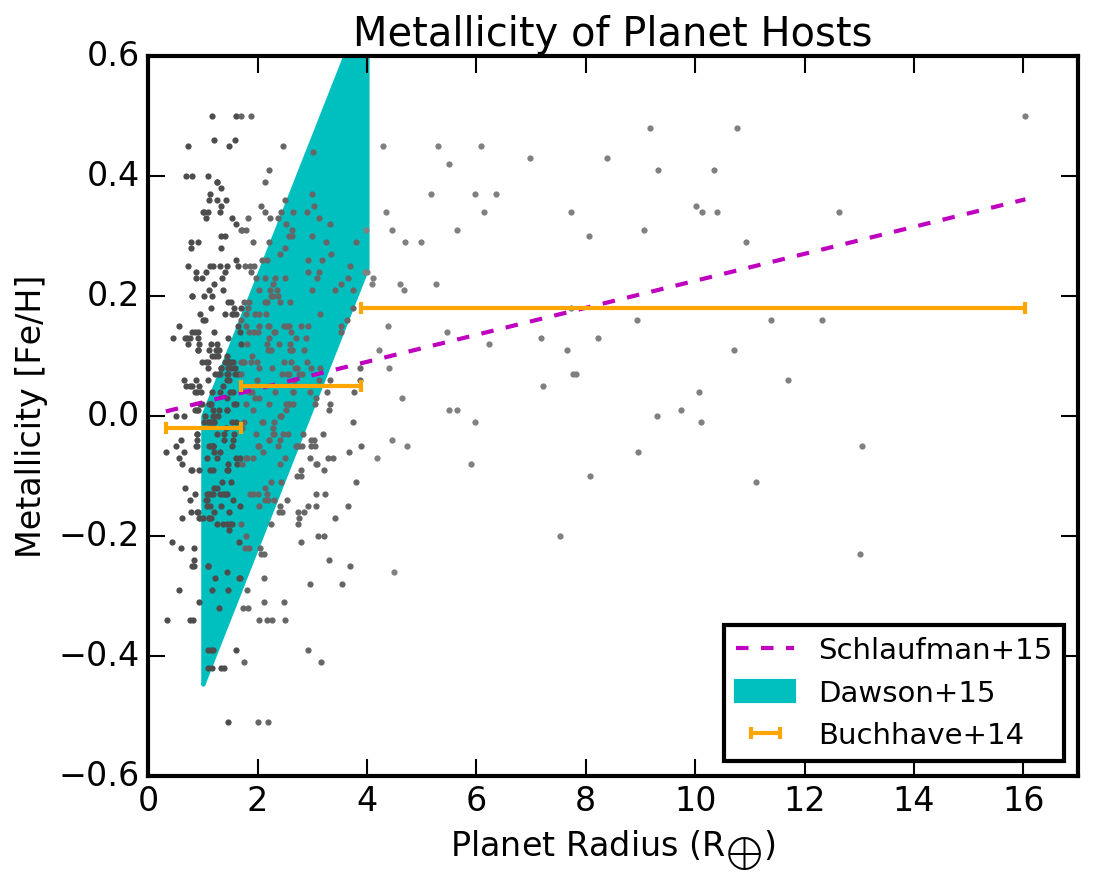}
\caption{Metallicity of planet host stars as function of planet radius. Points represent spectroscopic metallicities of \Kepler exoplanet hosts from \cite{2014Natur.509..593B}. The average host star metallicity correlates with planet radius, as indicated for a set of discrete radius bins shown in orange \citep{2014Natur.509..593B} and for a continuous planet radius-metallicity relation \citep{2015ApJ...799L..26S} shown with the dashed purple line. The expected range of planet radii from \insitu planet formation models by \cite{2015MNRAS.453.1471D} are shown in cyan.
}
\label{f:FeH:transit}
\end{figure}

\subsection{A Wide Range of Stellar Metallicity for sub-Neptunes}\label{s:Z:small}
Planets smaller than Neptune form around stars with a wide range of metallicities \citep{2008A&A...487..373S,2012Natur.486..375B}.
The planet-metallicity correlation identified for giant planets disappears when considering smaller planets (Fig \ref{f:FeH:transit}, \citealt{2014Natur.509..593B}).

\runinhead{Neptunes}
The first indications that Neptune-mass planets are not preferentially found around metal rich stars, as opposed to giant planet hosts, were found by \cite{2006A&A...447..361U} in a sample including M dwarfs planet hosts, and later confirmed by \cite{2008A&A...487..373S}. The possibility that a higher planet occurrence rate of Neptune-sized planets around M dwarfs contributed to this correlation was investigated by \cite{2010ApJ...720.1290G}, who recovered the wide range of stellar metallicities for Neptune-mass planet hosts in a sample of FGK dwarfs. This trend was confirmed by \cite{2011arXiv1109.2497M}, who show that planets less massive than $30$-$40 M_\oplus$ are equally common around metal-poor and metal-rich stars. The same metallicity-independence was found for M dwarfs hosting Neptune mass and smaller planets \citep{2012ApJ...748...93R,2013A&A...551A..36N}.

\runinhead{Transiting sub-Neptunes}
The large number of planets smaller than Neptune discovered by the \Kepler mission provide a unique opportunity to constrain the metallicity-dependence of planets down to Earth-sizes. Follow-up high resolution spectroscopy of \Kepler exoplanet hosts confirm that sub-Neptunes form around a wide range of stellar metallicities ($\FeH\approx[-0.6-0.5]$) and extend this trend to Earth-sized planets \citep{2012Natur.486..375B,2013ApJ...771..107E,2013ApJ...770...43M}.

\cite{2014Natur.509..593B} divided the sample into rocky planets ($R<1.7 R_\oplus$) and gas dwarfs ($1.7 R_\oplus < R < 3.9 R_\oplus$) and find that the mean metallicity of rocky planets is consistent with solar. On the other hand, the larger gas dwarfs have a mean metallicity of $\FeH=0.05$ that is significantly higher than non-planet hosting stars \citep{2015ApJ...808..187B}. Such a trend is consistent with 
a planet-metallicity correlation for the maximum size/mass of Neptunes \citep{2016MNRAS.461.1841C,2017AJ....153..142P}. However, \cite{2015ApJ...799L..26S} and \cite{2020AJ....160..253L} argue that the \Kepler data is better described by a continuous increase in metallicity with planet radius (Figure \ref{f:FeH:transit}), though \cite{2021AJ....162...69K} argue against such a relation.

Planet occurrence rates as a function of spectroscopic metallicity were calculated by \cite{2016AJ....152..187M} for a sample of 20,000 \Kepler target stars with medium resolution spectroscopy from \cite{2016A&A...594A..39F}. They find no difference in the occurrence rate of sub-Neptunes as a function of metallicity, except at orbital periods smaller than 10 days (see also \citealt{2018AJ....155...68W,2018AJ....155...89P}). This elevated occurrence rate at short orbital periods is consistent with the higher detection frequency of sub-Neptunes around metal-rich stars \citep{2015AJ....149...14W,2016ApJ...832..196Z}. 

Several other papers have pointed out trends in host star metallicity with the planet orbital period distribution \citep{2013ApJ...763...12B,2013A&A...560A..51A,2015MNRAS.453.1471D,2018PNAS..115..266D}, though there is some disagreement on the planet radius and orbital period where these transitions occur. 
The trend identified by \cite{2016OLEB...46..351A} that small ($<2~R_\oplus$) planets interior to the habitable zone may predominantly found in low-metallicity stars is tantalizing, but was found to not be significant when taking into account detection completeness by \cite{2016AJ....152..187M}.

\section{Trends With Stellar Mass}\label{s:M}
\begin{figure}
\includegraphics[width=\textwidth]{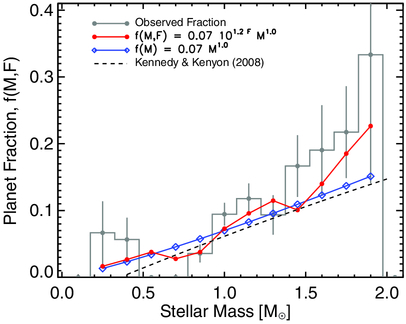}
\caption{Giant planet occurrence as function of stellar mass, from \cite{2010PASP..122..905J} figure 4.
The histogram shows the observed planet occurrence rate. The red red line show the predicted planet occurrence rate based on the metallicity distribution of stars in each stellar mass bin. The blue line shows the stellar-mass dependence at solar metallicity, compare to the predicted relation from the planet formation model by \cite{2008ApJ...673..502K}.
Figure reproduced from \cite{2010PASP..122..905J} with permission from the authors.
}
\label{f:M:RV}
\end{figure}

\begin{figure}
\includegraphics[width=\textwidth]{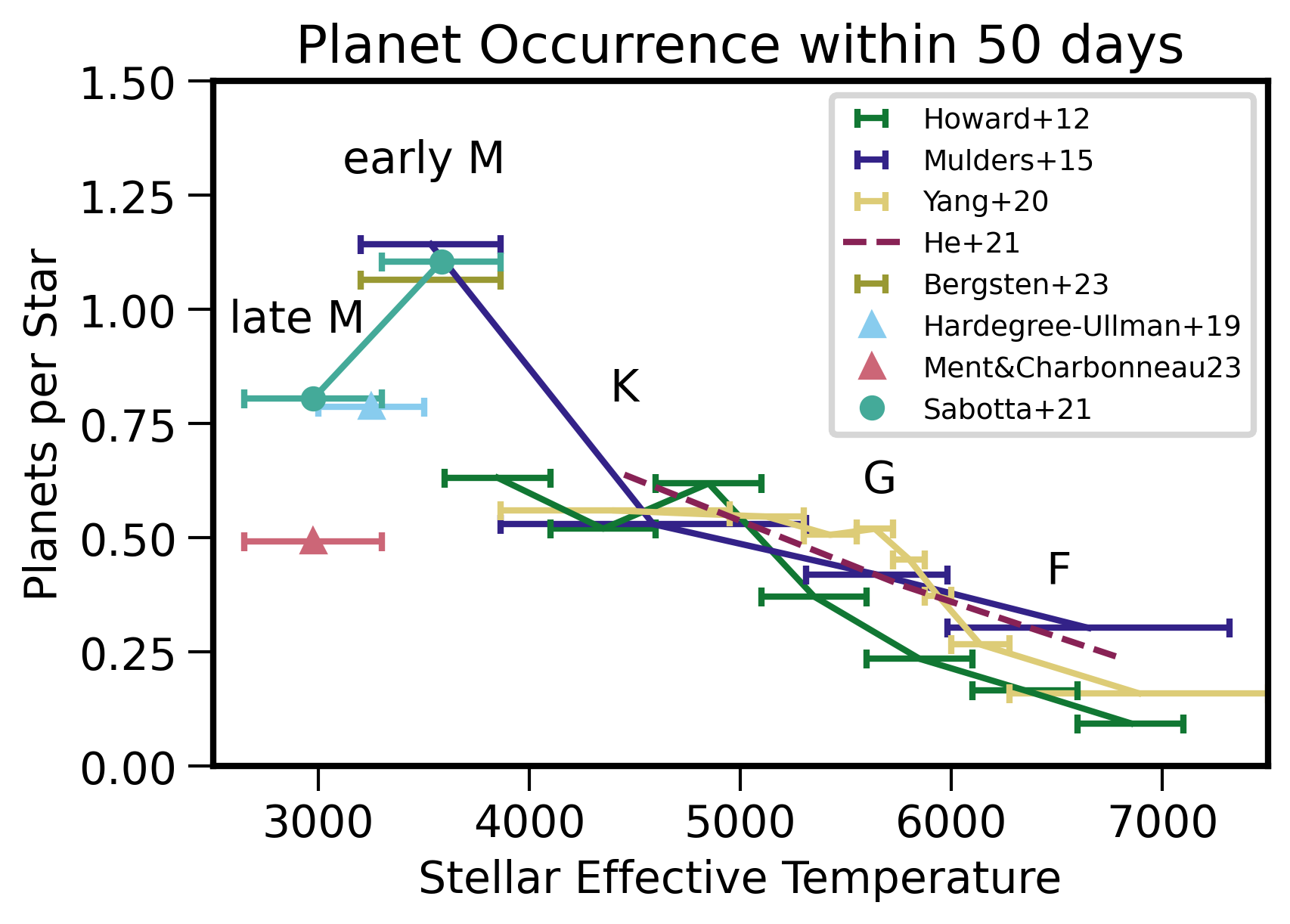}
\caption{Overview of planet occurrence rates as a function of effective temperature in the literature for planets between $1$-$4 R_\oplus$ and $P<50$ days. Occurrence rates were re-scaled assuming uniform occurrence in log period and log radius for purpose of this comparison. An increase in planet occurrence from spectral types F to early M is present across all studies. A break in this trends becomes visible towards late M dwarfs, though it should be noted those rates sample only short-period planets ($<10$ days) and are potentially lower limits to the occurrence out to $50$ days.
References -- \cite{2012ApJS..201...15H,2015ApJ...814..130M,2019AJ....158...75H,2020AJ....159..164Y,2021AJ....161...16H,2021A&A...653A.114S,2023AJ....165..265M,2023AJ....166..234B}.
}
\label{f:T:lit}
\end{figure}

The correlation between planet occurrence and stellar mass is dependent on planet size.  
Giant planets occur more frequently around higher-mass stars (Fig. \ref{f:M:RV}, \citealt{2010PASP..122..905J,2021ApJS..255...14F}), with a  linear dependence that is weaker than the quadratic dependence on metallicity. Sub-Neptunes, those found in abundance with the \Kepler survey, occur more frequently around low-mass M dwarfs (Fig \ref{f:T:lit}, \citealt{2015ApJ...814..130M,2021A&A...653A.114S}). 

\subsection{Giant planets}
Giant planets are found less frequently are low-mass M dwarfs than around sun-like stars and more frequently around evolved stars with higher masses (Fig. \ref{f:M:RV}). 
The giant planet occurrence scales roughly linear with stellar mass, $f_{\rm giant}\propto M_\star$, and is therefore weaker than the planet-metallicity correlation that scales quadratically as $f_{\rm giant}\propto \FeH^2$.

Tentative evidence for a decreased giant planet occurrence around M dwarfs compared to sun-like stars was found by \cite{2003AJ....125.2664L} and \cite{2006ApJ...649..436E}. The giant planet occurrence rate within 2.5 au increases by a factor of $\sim3$ from M stars to sun-like stars \citep{2006PASP..118.1685B,2008PASP..120..531C}. Planet occurrence rates for a sample of late K dwarfs support the positive correlation with stellar mass \citep{2013ApJ...771...18G}.

Taking metallicity into account, the giant planet occurrence rate increases roughly linear with stellar mass between M dwarfs, GK stars, and retired A stars (\citealt{2007ApJ...670..833J,2010PASP..122..905J,2018ApJ...860..109G}, Fig. \ref{f:M:RV}). The trend is also identified in radial velocity samples of MKGF stars \citep{2021ApJS..255...14F,2021A&A...653A.114S}.
The stellar-mass dependence has also been identified for giant planets at longer orbital periods by including radial-velocity trends and micro-lensing data  \citep{2014ApJ...781...28M,2014ApJ...791...91C}.
The planet occurrence rate around giant stars increases with stellar mass up to $\approx 2 M_\odot$ but decreases at larger stellar mass \citep{2015A&A...574A.116R,2016A&A...590A..38J,2022A&A...661A..63W}.

The giant planet occurrence rate in the \Kepler transit surveys is low, consistent with the predictions from radial velocity surveys \citep{2013ApJ...767...95D}. The occurrence rate of giant planets with orbital periods less than 50 days is more than two times higher for FGK stars than M stars \citep{2015ApJ...814..130M}. Using TESS data, \cite{2023AJ....165...17G} and \cite{2023MNRAS.521.3663B} show that he occurrence of transiting giant planets decreases with stellar mass from G dwarfs to early M dwarfs to late M dwarfs. Somewhat surprisingly, the hot Jupiter occurrence rate from TESS also falls towards F and A stars \citep{2022MNRAS.516...75B,2023AJ....165...17G}. This suggests a peak Hot Jupiter occurrence around one solar mass, in contrast to the colder giant planets from radial velocity surveys peaking around two solar masses \citep[e.g.][]{2022A&A...661A..63W}.

\subsection{The Sub-Neptune Exoplanet Population}
Planet smaller than Neptune far outnumber their larger counterparts \citep[e.g.][]{2010Sci...330..653H,2012ApJS..201...15H}. These sub-Neptunes are show a different dependence on host star mass than giant planets: they become more frequent towards lower mass stars (Fig. \ref{f:T:lit}). 
Neptune-mass and smaller planets are commonly found around M dwarfs in radial velocity surveys, where the smaller mass ratio between star and planet favors planet detection compared to FGK stars. The sub-Neptune exoplanet population is most constrained by \Kepler, whose detection efficiency reaches down to earth radii and smaller at short orbital periods, supported by \TESS and radial velocity surveys.

\runinhead{High planet occurrence around low-mass stars}
The increase in exoplanet occurrence with decreasing effective temperature (a proxy of stellar mass) was discovered by \cite{2012ApJS..201...15H}, see also Figure \ref{f:T:lit}. Taking into account differences in detectability between stars of different sizes in the \Kepler survey, they find that the occurrence rate of planets between 2-4 $R_\oplus$ is anti-correlated with effective temperature and increases by a factor 7 between the hottest stars in the sample (late F stars) and the coolest stars (early M dwarfs). This trend was extended down to Earth-sized planets by \cite{2015ApJ...798..112M}, who found an increase in planet occurrence rate between F,G,K, and M type stars at all orbital periods. The occurrence rate of (early) M stars compared to FGK stars is a factor $\sim2$-$4$ higher at planet radii between $1$ and $\sim 3 R_\oplus$ \citep{2015ApJ...814..130M,2016MNRAS.457.2877G,2023AJ....165..262Z}. Radial velocity surveys show a similar increase in the planet occurrence rate around M dwarfs \cite{2021A&A...653A.114S}. Taking into account system architectures, \cite{2020AJ....159..164Y} and \cite{2021AJ....161...16H} show it is the fraction of stars with planetary systems that increases towards lower mass stars, while the number of planets per planetary system remains roughly constant. 

Figure \ref{f:T:lit} shows the occurrence of rate of sub-Neptunes ($1-4~R_\oplus$) at orbital periods less than 50 days as a function of stellar effective temperature as estimated by different studies. For purposes of this comparison, occurrence rates were rescaled when only estimates for a different range of planet properties were available, assuming a uniform occurrence in log planet radius and log orbital period. While there is significant scatter in occurrence rates at similar effective temperatures, the elevated planet occurrence rates around M dwarfs compared to FGK stars is clearly present, as well as the more gradual decrease in planet occurrence rate within the FGK stars.

\begin{figure}
\includegraphics[width=0.5\textwidth]{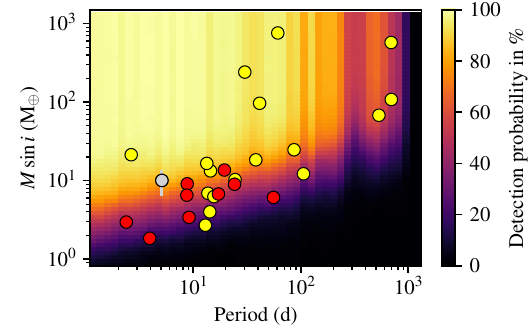}
\includegraphics[width=0.5\textwidth]{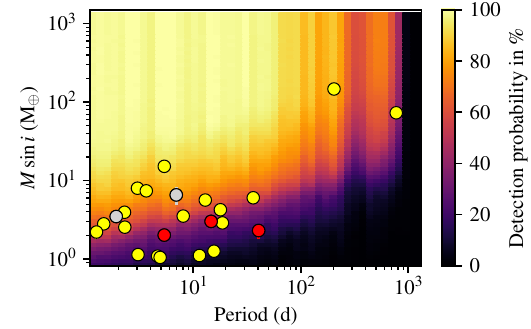}
\put(-280,105){$M_\star > 0.34~M_\odot$}
\put(-110,105){$M_\star < 0.34~M_\odot$}
\caption{Radial velocity detected exoplanets around higher-mass (left) and lower mass (right) M dwarfs from the CARMENES survey \citep{2021A&A...653A.114S,2023A&A...670A.139R}. Aside from a decrease in the number of giant planets, sub-Neptunes tend to be smaller and located closer to the star for lower-mass stars. 
Figure based on \cite[][Fig. 3]{2021A&A...653A.114S}  with updated data as in \cite[][Fig. 10]{2023A&A...670A.139R}.
}
\label{f:carmenes}
\end{figure}

\runinhead{Trends within M dwarfs}
M dwarfs are not a homogenous group of stars, but span a large range of stellar masses, radii, and luminosities. Exoplanet properties also vary significantly across this range (see Fig. \ref{f:carmenes}), though it can be hard to characterize the population around the lowest mass M dwarfs due to a lack of stars sufficiently bright for exoplanet detection. 
At the low-mass end of its stellar population, the \Kepler survey was mainly sensitive to early M and late K dwarfs \citep{2015ApJ...807...45D,2023AJ....166..234B}. A small number of even lower-mass mid M dwarfs where also observed, with a planet occurrence rate at orbital periods less than 10 days consistent with that of the early M dwarfs \citep{2019AJ....158...75H}. \TESS, being an all sky survey, has observed a larger number of M dwarfs, including many lower mass mid and late M dwarfs.  Focusing on the nearest and most-well characterized mid-to-late M dwarfs, \cite{2023AJ....165..265M} find a high occurrence ($\approx 60\%$) of super-earth sized planets ($<1.5~R_\oplus$) within an orbital period of 7 days. However, the authors find a lack of mini-Neptunes ($> 1.5~R_\oplus$), presenting a break with the ever-increasing sub-Neptune occurrence observed in F, G, K, and early M dwarfs (see Fig. \ref{f:T:lit}).

A similar trend is seen in the CARMENES radial velocity survey of M dwarfs (Fig. \ref{f:carmenes}, \citealt{2021A&A...653A.114S}). Around the more massive M dwarfs, sub-Neptunes have masses between $2$ and $20~M_\oplus$ and are mostly located at orbital periods between $10-100$ days (left panel of Fig. \ref{f:carmenes}), coinciding with regions of high planet occurrence of transiting sub-Neptunes around early M and FGK stars \citep[e.g.][]{2015ApJ...798..112M}. Around less massive M dwarfs ($< 0.35~M_\odot$) planets are on average less massive ($1-10~M_\oplus$) and located mostly within an orbital period of 10 days (right panel of Fig. \ref{f:carmenes}). 

As such, the planet population around late M dwarfs starts to resemble the archetype of the Trappist-1 system \citep{2017Natur.542..456G}, with many (super)Earth-sized planets at short orbital periods, but lacking the larger mini-Neptunes at somewhat longer periods so prominent around early M dwarfs and FGK stars.

\section{Constraints On Planet Formation Mechanisms}
The dependence of the exoplanet population on host star properties provides constraints on planet formation mechanisms. The positive correlations of giant planet occurrence rate with stellar mass and metallicity support the core accretion scenario of giant planet formation. The constraints provided by the lack of a clear correlation for sub-Neptunes with stellar metallicity and the anti-correlation with stellar mass have yet to be determined, but point to the role of \pebbles in the planet formation process. These trends indicate that planet formation is a robust and efficient process that takes place in a variety of environments. 

\subsection{Formation of Giants Planets}
The core accretion scenario postulates that giant planets form ``bottom up'' with the formation of a $\sim 10 M_\oplus$ solid core followed by a subsequent phase where most of the gas is accreted \citep{1996Icar..124...62P}. 
As the envelope has to be accreted before the protoplanetary disk gas is dispersed, typically $\sim 3$ million years \citep[e.g][]{2009AIPC.1158....3M}, the growth of the core has to be sufficiently rapid to allow giant planets to form. Pebble accretion \citep{2010A&A...520A..43O,2012A&A...544A..32L} is often invoked to facilitate this rapid growth of the core, see \cite{2023ASPC..534..717D} for an extended review.  The time scale for core growth depends on the amount of material locally available in the disk, i.e. the solid surface density. Giant planets thus form only in protoplanetary disks with a sufficiently high surface density of solids \citep[e.g.][]{2000ApJ...537.1013I,2002ApJ...581..666K}.

The stellar metallicity is a tracer of the solid inventory in protoplanetary disks at the onset of planet formation. Stars and protoplanetary disks inherit the same metallicity from the parental molecular cloud. Stars with a high metallicity formed with disks with a high solid surface density, and are therefore more likely to form giant planets. Numerical simulations of core formation and envelope accretion in disk with different metallicities consistently reproduce the observed giant planet-metallicity correlation \citep[e.g.][]{2004ApJ...616..567I,2005A&A...430.1133K,2008ApJ...673..487I,2009A&A...501.1161M}. 

A similar argument can be made for the dependence of the giant exoplanet population on stellar mass.
Protoplanetary disks mass, both gas and solids, scales with stellar mass (see Fig. \ref{f:M:disk}), while giant planets are more likely to form in more massive disks \citep[e.g.][]{2008Sci...321..814T}. 
By extension, the core accretion model predicts a positive correlation between giant planet occurrence and stellar mass.
Based on analytical estimates, \cite{2004ApJ...612L..73L} predict fewer giant planets around M dwarfs. Detailed numerical simulations show a nearly linear dependence of giant planet occurrence on stellar mass \citep{2005ApJ...626.1045I,2008ApJ...673..502K,2011A&A...526A..63A}, consistent with the observed trends (Figure \ref{f:M:RV}, however, see \citealt{2022A&A...664A.180S}).

\runinhead{Gravitational instability}
In the gravitational instability scenario, giant planets form ``top down'' from the contracting gas in massive protoplanetary disks \citep{1997Sci...276.1836B}. This formation mechanism predicts different dependence on stellar mass and metallicity. A high disk metallicity inhibits cooling and contraction of the gaseous envelope, and therefore giant planets should form more efficiently around low-metallicity stars \citep{2010MNRAS.406.2279M}. Gravitational instabilities should also form planets efficiently around M dwarfs \citep{2006ApJ...643..501B}. The observed positive correlations between giant planet occurrence with stellar mass and metallicity indicate that planets at short orbital periods likely did not form through gravitational instability in a protoplanetary disks.

\runinhead{Increasing Stellar Metallicity by Accretion of Planets} 
Accretion of planets can increase the stellar metallicity if planets are more metal-rich than their host star. 
It was initially suggested that the enhanced metallicity of planet-hosting stars is caused by the accretion of planets or solids \citep{1997MNRAS.285..403G}, instead of planet formation being more efficient around more metal-rich stars. 
The observational signature of planetary accretion is only large enough if the accreted metals are not mixed throughout the entire star, but remain near the surface in the convective zone. 
In F and A stars, the convective zone is thin enough that the accretion of solids can lead to a metallicity increase that is consistent with observations \citep{1997ApJ...491L..51L}. 
For lower-mass stars the convective zones are deeper and the metallicity signature of accreted planets should drop below detectable levels for G type and earlier stars \citep{1997ApJ...491L..51L}. This prediction is inconsistent with the observed giant-planet metallicity relation for these stars \citep[e.g.][]{2005ApJ...622.1102F} as well as for M dwarfs \citep[e.g.][]{2013A&A...551A..36N}. Once stars evolve off the main-sequence, mixing should increase, thereby diluting the metallicity enhancement from planetary accretion. However, the planet metallicity correlation is also observed in evolved stars \citep[e.g.][]{2010PASP..122..905J,2015A&A...574A.116R,2016A&A...590A..38J}.
Hence, the hypothesis that planetary accretion causes the planet-metallicity correlation is no longer supported by observational evidence.

\subsection{Formation of sub-Neptunes}
The different scaling laws with stellar mass and metallicity indicate a different formation history for giant planets and sub-Neptunes. Indeed, the comparison between the predictions of the core accretion model \citep{2008ApJ...685..584I,2009A&A...501.1139M} with the population of sub-Neptunes detected in radial velocity surveys \citep{2010Sci...330..653H} and the \Kepler transit survey \citep{2012ApJS..201...15H} show that the predicted ``planet desert'' at orbital period less than 50 days is indeed well-populated, highlighting the need to amend planet formation theory for sub-Neptunes \citep[e.g.][]{2021A&A...656A..70E,2019ApJ...887..157M}.

The moniker of `core accretion' is not particularly useful when discussing sub-Neptunes as they are, almost by definition, the planets that did not accrete massive gaseous envelopes. The planet formation mechanisms discussed here are almost exclusively focused on sub-Neptunes and it should be kept in mind that these new mechanisms are to amend, not replace, core accretion theory.

Several planet formation mechanisms have been proposed to explain the presence of small planets at short orbital periods \citep[e.g.][]{2008MNRAS.384..663R}. The two mechanisms that are of most relevance here are \insitu formation and \migration . In addition, \pebbles, combined with inward radial drift, is increasingly invoked as a mechanism to form close-in sub-Neptunes (see \citealt{2023ASPC..534..717D} for a review).

\runinhead{In Situ Formation}
The \insitu Formation scenario for exoplanets is based on terrestrial planet formation in the Solar System.
Planetary embryos in the protoplanetary disk can grow through oligarchic growth to a fraction of the final planet mass, typically Mars-size at 1 au \citep[e.g.][]{1987Icar...69..249L,2000Icar..143...15K,1998Icar..131..171K}
After the gas disk disperses, gravitational interactions increase the protoplanet eccentricities and makes them collide and merge, leading to a phase of giant impacts in which planets grow to their final masses \citep[e.g.][]{1998Icar..136..304C,1985Sci...228..877W}. As the majority of the accreted material is sourced from a region close to the planets final orbit, the planet mass is directly dependent on the local surface density of planetary building blocks \citep{2002ApJ...581..666K}. 
\cite{2013MNRAS.431.3444C} proposed that planetary systems observed with \Kepler could have formed \insitu in disks that are on average more massive than the protoplanetary disk around the sun. N-body simulations of the giant impact phase show that disks with high surface density of solids in the inner regions can indeed form Kepler-like planetary systems \citep{2012ApJ...751..158H,2013ApJ...775...53H,2020ApJ...891...20M,2020ApJ...897...72M}. 

\runinhead{Pebble Accretion}
The \pebbles hypothesis is based on the rapid growth of planetary embryos through aerodynamically assisted accretion \citep{2010A&A...520A..43O,2012A&A...544A..32L}. Combined with inward radial drift of solids which brings a large amount of pebbles into the inner disk, super-earths can directly form \citep[e.g.][]{2019A&A...627A..83L}. When planetary cores grow massive enough to perturb the surrounding gas, a pressure bump can form that limits the subsequent growth by pebbles. The mass where this happens is called the pebble isolation mass \citep{2014A&A...572A.107L}, which in typically in the super-earth or sub-Neptune regime \citep{2019A&A...632A...7L}.

\runinhead{Planet Migration}
The Planet Migration hypothesis is built on he theoretical expectation that low-mass planets embedded in a gaseous disk undergo rapid inward migration (Type-I migration, \citealt{1997Icar..126..261W}). Because planetary embryos can grow to larger sizes in the outer disk where more material is available, \migration does not require disks to be particularly massive \citep[e.g.][]{2013ApJ...764..105S}. The type-I migration time scales are short ($<10^5$ years) compared to the disk life time of a few million years \citep[e.g.][]{2009AIPC.1158....3M}, and migration needs to be halted in the inner disk. Possible mechanisms to stall migration include an inner disk cavity \citep[e.g.][]{2007ApJ...654.1110T}, resonant capture by other planets, and regions of outward migration due to disk density and temperature structure \citep{2014A&A...567A.121D,2014A&A...569A..56C}. The largest challenge for planet migration hypothesis is that the observed multi-planet systems are often not in  
orbital resonances as predicted from convergent migration \citep{2014ApJ...790..146F}, though different mechanisms have been proposed to break resonances after formation \citep{2012MNRAS.427L..21R,2014AJ....147...32G,2017MNRAS.470.1750I}.

These three scenarios are not mutually exclusive. \migration models often include a growth phase including pebble accretion, and a giant impact phase during or after migration \citep[e.g.][]{2021A&A...650A.152I}. \insitu formation models often invoke, explicitly or implicitly, an inward migration phase of solids to increase the amount of planetary building blocks in the inner disk \citep[e.g.][]{2012ApJ...751..158H}, similar to the radial drift of pebbles \citep{2015ApJ...809...94M,2020A&A...642A..75V}.
Despite these nuances, \migration, \insitu and \pebbles remain useful concepts in discussing the origin the observed trends with stellar mass and metallicity.

\runinhead{Metallicity Dependence}
The stellar metallicity is a direct measure of the amount of condensible solids that was available for planet formation in the disk. 
The base expectation is that the mass in planetary systems correlates positively with disk metallicity. 
The \insitu formation simulations in \cite{2015MNRAS.453.1471D} show that, for a range of metallicity of a factor 10, the predicted planet radii vary between $1$-$4 R_\oplus$, with significant scatter (see also Figure \ref{f:FeH:transit}). A clear planet size-metallicity relation is not seen in the observed population of small exoplanets. There is tentative evidence for a lack of rocky ($<2 R_\oplus$) planets at high metallicities at a limited orbital period range \citep{2015MNRAS.453.1471D,2016OLEB...46..351A}, though this trend may not be statistically significant when taking into account survey completeness \citep{2016AJ....152..187M}. The predicted lack of sub-Neptunes ($2-4 R_\oplus$) at low metallicity is not observed. 
However, a planet size-metallicity relation appears to be present for planets more massive than Neptune \citep{2016MNRAS.461.1841C,2017AJ....153..142P}. The elevated host star metallicity of transiting sub-Neptunes \citep{2014Natur.509..593B,2015ApJ...808..187B} seems to support \insitu formation scenario, perhaps with a much wider range in planet radii than predicted by \citep{2015MNRAS.453.1471D}, originating from a wide range in disk masses \citep{2021AJ....162...69K}. On the other hand, the planet size-metallicity relation inferred for \Kepler planets by \cite{2015ApJ...799L..26S} is significantly shallower than linear, see Figure \ref{f:FeH:transit}. Hence, it is clear that the planet-metallicity correlation predicted by \insitu formation models is not observed. 

In the \migration scenario, the mass of planets that form in outer disk is also dependent on disk metallicity. However, the subsequent inward migration may shape the observed distribution of exoplanets in a different way. \cite{2014A&A...569A..56C} find that super-earths consistently form in a set of simulations varying the dust-to-gas ratio, a proxy of metallicity, by a factor 4. The total mass of planetary systems show an almost linear dependence on metallicity, and hence does not deviate significantly from the predictions of \insitu formation models. \cite{2016MNRAS.457.2480C} model the growth and migration of super-earths, and find that planets in low-metallicity disks do not reach the mass required for efficient inward migration, and hence close-in super-earths do not form. Instead small mobile bodies (pebbles) must play an important role in the formation of super-earths, though the predictions of such a model have not been explored in detail.

In the \pebbles scenario, the pebble isolation mass could potentially play a role at erasing any metallicity dependence. Because this mass scales with gas disk properties, and not with the amount of solids available, an increased metallicity does not directly affect planet size, limiting  the role of metallicity in determining final planet properties. 

\begin{figure}
\includegraphics[width=\textwidth]{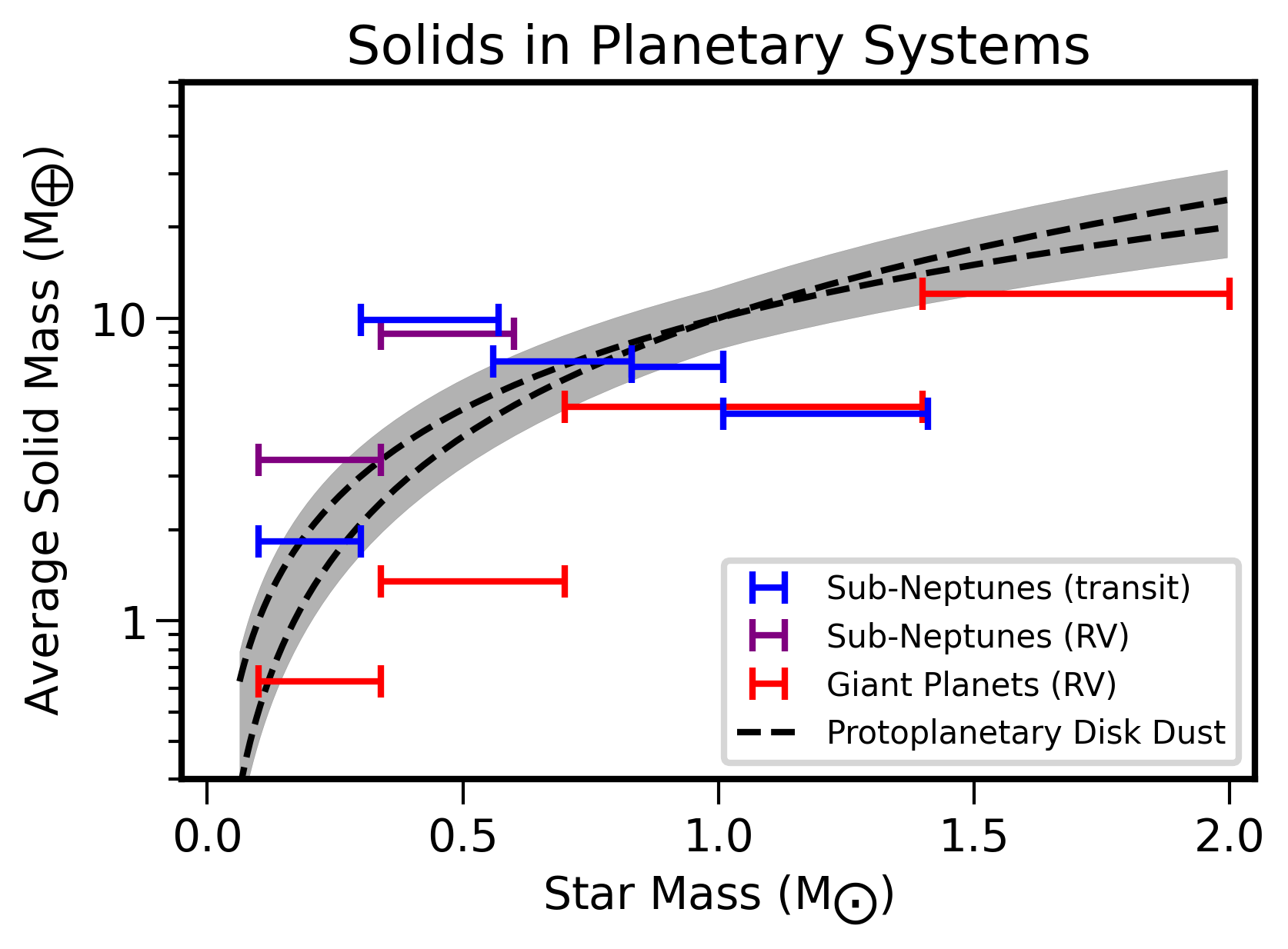}
\caption{Average solid mass of planetary systems around stars of different masses. 
This quantity is calculated by multiplying the planet occurrence rate by an estimate of the planet mass, see text for details. 
Sub-Neptunes at short orbital periods (blue, purple) are compared to giant planets out to a few au (red) and the median dust masses of Class II protoplanetary disks in the Chamaeleon I star forming region, where the two different slopes of the two lines reflect some of the uncertainties in the derived stellar-mass dependence. An excess of solids in sub-Neptunes around M dwarfs is evident, posing a riddle for planet formation models.
}
\label{f:M:disk}
\end{figure}

\runinhead{Low Mass Stars}
The anti-correlation between the occurrence of planets at short orbital period and the stellar mass poses an urgent problem for planet formation theories: How to explain the elevated planet occurrence rates of low mass stars if less material is available in their disks to form planets? Figure \ref{f:M:disk} illustrates this issue, showing the estimated amounts of solids around stars of different masses in:
\begin{itemize}
\item Planets at orbital periods less than 150 days around F, G, K, and early M dwarfs in the \Kepler survey, from \cite{2015ApJ...814..130M}, estimated assuming the mass-radius relation from \cite{2016ApJ...825...19W} and a core mass of $20 ~M_\oplus$ per giant planet. 
\item Transiting planets around mid-to-late M dwarfs observed with TESS from \cite{2023AJ....165..265M}, assuming an average planet mass of $3 ~M_\oplus$.
\item Radial velocity detected sub-Neptunes around M dwarfs from \cite{2021A&A...653A.114S,2023A&A...670A.139R}, using an average mass of $3 ~M_\oplus$ for late M dwarfs and $6 ~M_\oplus$ for early M dwarfs.
\item Giant planets out to 2.5 au around GK stars and retired AF stars from \cite{2010PASP..122..905J}, assuming $60 M_\oplus$ of solids per Jupiter-mass giant planet \citep{2016ApJ...831...64T}.
\item Giant planets around M dwarfs from \cite{2023A&A...670A.139R} assuming $30 ~M_\oplus$ of solids per giant planet.
\item Protoplanetary disks in the Chamaeleon I star forming region, from \cite{2016ApJ...831..125P}, probing solids at scales of $\sim 10-100$ au.
\end{itemize}
While the solids in giant planets and protoplanetary disks show a positive scaling with stellar mass, this relation breaks down for sub-Neptunes at short orbital periods.
The estimated amount of solids in M dwarf planetary systems is higher than that in sun-like stars \citep{2015ApJ...814..130M,2017MNRAS.470L...1G}, reflecting the trend in planet occurrence.

\begin{figure}
\includegraphics[width=\textwidth]{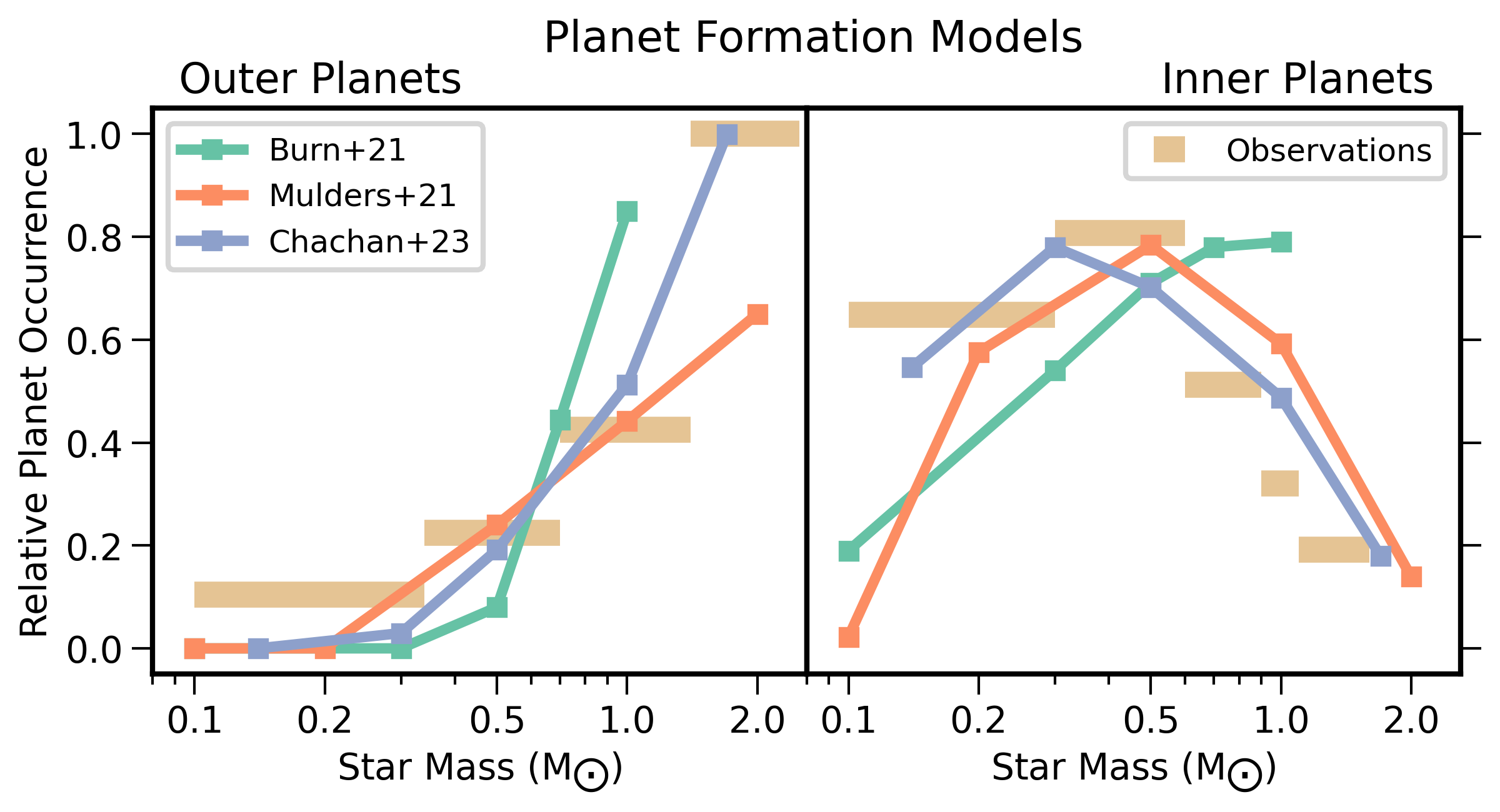}
\caption{
Relative planet occurrence as function of stellar mass predicted by different planet formation models. The green line shows the Bern planet population synthesis model \citep{2021A&A...656A..72B} based on planetesimal accretion and planet migration. The orange line \citep{2021ApJ...920L...1M} and purple line \citep{2023ApJ...952L..20C} show pebble accretion models without migration. While all models reproduce the increase in giant planet occurrence rate with stellar mass (left panel), only the pebble accretion models show the observed decrease in the occurrence of sub-Neptunes from late M to F stars.
}
\label{f:pfmodels}
\end{figure}

A strict \insitu formation model for sub-Neptunes, where planet mass is directly related to mass available in disk \citep[e.g.][]{2007ApJ...669..606R,2015ApJ...804....9C}, is not favored by stellar-mass dependencies as evident from Figure \ref{f:M:disk}. Models that include \migration or \pebbles move material around radially in the disk, leading to different efficiencies of planet formation at different stellar masses. Figure \ref{f:pfmodels} shows the predictions planet occurrence rates of inner sub-Neptunes and outer giant planets as function of stellar mass.

The planet population synthesis model using planetesimal accretion and planet migration from \cite{2021A&A...656A..72B}
 show positive correlation between planet occurrence and stellar mass for both sub-Neptunes and Giant planets, the former inconsistent with observed trends. Two \pebbles models better explain the occurrence of exoplamets as function of stellar mass.  \citep{2021ApJ...920L...1M} use a  growing giant planets outside the snow line to block the flow of pebbles into the inner disk, suppressing the growth of super-earths. Because giant planet form more efficiently around the more massive stars, the occurrence of sub-Neptunes is suppressed more for more massive stars, explaining the observed downturn in planet occurrence towards higher stellar mass. In the model of \citep{2023ApJ...952L..20C}, planet formation in the inner disk through pebble accretion is most efficient around stars between $0.3-0.5~M_\odot$, predicting a peak in the planet occurrence for early M dwarfs and a lower occurrence for more massive stars (See fig. \ref{f:pfmodels}).

Both pebble accretion models from predict the observed downturn in planet occurrence toward late M dwarfs \citep{2021A&A...653A.114S,2023AJ....165..265M}, though it should be noted that that the model from \cite{2021A&A...656A..72B} predicts a similar  scaling relation in this stellar mass range. These pebble accretion models also predict the observed decrease in planet mass for late M dwarfs (see also \citealt{2019A&A...632A...7L}).

\section{Conclusions}
Giant planets occur more frequent around more massive and more metal-rich stars. These trends support the core-accretion scenario for giant planet formation in which accretion of a gaseous envelope starts after a sufficiently rapid assembly of a massive rocky core. The threshold for reaching the critical core mass is reached more easily in protoplanetary disks with a larger amount of condensible solids around metal-rich stars and in more massive disks around more massive stars.

These results stand in contrast to the population of exoplanets smaller than Neptune, those that are found in abundance with \Kepler and \TESS and represent the bulk of the exoplanet population. These sub-Neptunes are found around stars with a wide range of metallicities, indicating that planet formation is a robust process that occurs efficiently in a variety of environments. Curiously, sub-Neptunes occur much more frequently around low-mass M dwarfs than around solar-mass stars, though this trend reverses toward the lowest-mass stars, late M dwarfs. Pebble drift and accretion likely plays a key role in boosting the planet formation efficiency for M dwarfs, explaining the abundance of exoplanets around low-mass stars.

\begin{acknowledgement}
I would like to thank Ignasi Ribas and Silvia Sabotta for providing Figure \ref{f:carmenes}.
G.D.M. acknowledges support from FONDECYT project 11221206, from ANID --- Millennium Science Initiative --- ICN12\_009, and the ANID BASAL project FB210003. The results reported herein benefitted from collaborations and/or information exchange within NASA’s Nexus for Exoplanet System Science (NExSS) research coordination network sponsored by NASA’s Science Mission Directorate and project “Alien Earths” funded under Agreement No. 80NSSC21K0593
\end{acknowledgement}

\bibliographystyle{spbasicHBexo}  
\bibliography{update}

\end{document}